\def\ArtDir{main/}
\documentclass{ws-ijgmmp}
\def\ArtDir{main/}
\begin{document}

\catchline{}{}{}{}{}

\title{Foundational issues in $f(T)$ gravity theory}

\author{Alexey Golovnev}

\address{Centre for Theoretical Physics, British University in Egypt, 11837 El Sherouk, Cairo, Egypt \email{agolovnev@yandex.ru}}

\author{Mar\'ia-Jos\'e Guzm\'an}

\address{Departamento de F\'isica y Astronom\'ia, Facultad de Ciencias, Universidad de La Serena,\\
Av. Juan Cisternas 1200, 1720236 La Serena, Chile}\address{
Laboratory of Theoretical Physics, Institute of Physics, University of Tartu,\\ 
W. Ostwaldi 1, 50411 Tartu, Estonia\\
\email{maria.j.guzman.m@gmail.com}}

\maketitle

\begin{abstract}
We give a short review on the status of research on the theoretical foundations of $f(\mathbb T)$ gravity theories. We discuss recent results on perturbative and non-perturbative approaches, causality and degrees of freedom, and discuss future directions to follow.

\end{abstract}


\section{Introduction}

Teleparallel gravity and its modifications represent a description of the gravitational interaction via the torsion instead of the curvature of spacetime \cite{Aldrovandi:2013wha,Golovnev:2018red}. The main starting point is, given a metric on the spacetime manifold, to weigh  connections which differ from the Levi-Civita one. The teleparallel description is given in terms of tetrads $e^{a}_{\mu}$ instead of the metric itself which is retrieved by the relation
\begin{equation}
g_{\mu\nu}=\eta_{ab}e^a_{\mu}e^b_{\nu},
\end{equation}
however it requires absolute parallelism of spacetime, which is not so restrictive though in causal $3+1$ geometry as it might have been if we were in a different dimension.

Historically, the teleparallel gravity was based on the Weitzenb{\" o}ck connection. It is defined as (we distinguish the inverse tetrad by using the capital letter $E$)
\begin{equation}
\label{conn}
\Gamma^{\alpha}{}_{\mu\nu}=e_a^{\alpha}\partial_{\mu}E^a_{\nu},  
\end{equation}
and one can check by direct calculation that the Riemann curvature $R^{\rho}{}_{\mu\nu\lambda}$ is zero. Note that $\Gamma^{\alpha}{}_{\mu\nu}$ does not transform correctly (``correctly'' would mean no transformation at all) under local Lorentz rotations of tetrads, and therefore it has a non-trivial dependence not only on the metric, but also on the choice of the reference frame. We will come back to this discussion later. Now let us remind that on top of zero curvature this connection also has zero non-metricity but is of course characterised by a non-trivial torsion tensor $T^{\alpha}_{\hphantom{\alpha}\mu\nu}=\Gamma^{\alpha}{}_{\mu\nu}-\Gamma^{\alpha}{}_{\nu\mu}$. 
 
Any metric connection on a manifold is different from the Levi Civita $\mathop{\Gamma^{\beta}{}_{\mu\nu}}\limits^{(0)}$ one by a tensorial quantity (as a difference of two connections)
\begin{equation}
K_{\alpha\mu\nu} = g_{\alpha\beta} \left(\Gamma^{\beta}{}_{\mu\nu} - \mathop{\Gamma^{\beta}{}_{\mu\nu}}\limits^{(0)}\right)=\frac12 \left(T_{\alpha\mu\nu}+T_{\nu\alpha\mu}+T_{\mu\alpha\nu}\right)
\end{equation}
called contortion tensor. Using it one can easily find the relation between two curvature tensors, and after contraction of indices this relation gives
\begin{equation}
\label{telepid}
0=R(\Gamma)=R(\mathop\Gamma\limits^{(0)})+2 \mathop\bigtriangledown\limits^{(0)}{\vphantom{\bigtriangledown}}_{\mu}T^{\mu}+ \mathbb T
\end{equation}
where we use the torsion vector $T_{\mu}= T^{\alpha}_{\hphantom{\alpha}\mu\alpha}$ and the torsion scalar ${\mathbb T}= T_{\alpha\mu\nu}S^{\alpha\mu\nu}$ with the superpotential tensor
\begin{equation}
S_{\alpha\mu\nu}=\frac12\left(K_{\mu\alpha\nu}+g_{\alpha\mu}T_{\nu}-g_{\alpha\nu}T_{\mu} \right)
\end{equation}
which has the same antisymmetry in the last two indices as the torsion tensor (note that in many works the factor of one half is moved from this definition of the superpotential to the definition of the torsion scalar in terms of the superpotential).

We see that the torsion scalar of Weitzenb{\"o}ck connection is different from the Levi Civita scalar curvature only by a surface term. It explains why the Teleparallel Equivalent of General Relativity (TEGR) with $\mathbb T$ as its Lagrangian density is equivalent to the Einstein-Hilbert action defining GR. However modifications such as $f({\mathbb T})$ \cite{Ferraro:2006jd,Bengochea:2008gz} and others are genuinely new models.

\section{Lorentz invariance}

As we mentioned above, the chosen connection (\ref{conn}) does not respect the local Lorentz symmetry. It is not a problem of the particular functional form, but a consequence of prescribing some particular dependence of the connection coefficients on the tetrad and nothing else. If this dependence is not restricted to only metric like for the Levi Civita connection then one needs a spin connection to make the picture locally Lorentz invariant. 

The spin connection would be taking into account the effects of changing the reference frame. Indeed, a choice of frame can be described by choosing the tetrad for a given metric. However, the pure tetrad formulation of $f(T)$ does not allow for free choice like this, since the local Lorentz rotations are not a symmetry of the model. If a solution of the equations of motion is found, one cannot make an arbitrary local Lorentz transformation of the tetrad and expect that the new tetrad satisfies the equations of motion, because local Lorentz invariance is lost in the model. However, adding a proper non-zero spin connection (in the way defined below) makes the tetrad a solution of the equations of motion again, and it becomes even the same solution in terms of both the metric and the torsion tensor.

It is well known how to make the $f(T)$ gravity locally Lorentz invariant \cite{Aldrovandi:2013wha,Krssak:2015oua,Golovnev:2017dox,Hohmann:2018rwf,Krssak:2018ywd}. The basic meaning is that the Weitzenb{\" o}ck connection is treated then as given in one particular frame, and formula (\ref{conn}) is supplemented by spin connection $\omega^{a}{}_{b\mu}$
\begin{equation} \label{fullsp}
\Gamma^{\alpha}{}_{\mu\nu}=e_a^{\alpha}(\partial_{\mu}E^a_{\nu}+\omega^{a}{}_{b\mu}E^b_{\nu})  
\end{equation}
which also transforms when the Lorentz group is acting, and in particular it becomes non-zero if a local Lorentz transformation is applied to the frame where it was zero.

To be more precise, in case of the Lorentz rotation $E^a_{\mu}\rightarrow \Lambda^a_b E^b_{\mu}$ the spin connection transforms as
$$\omega^a_{\hphantom{a}b\mu}\rightarrow \Lambda^a_c \omega^c_{\hphantom{a}d\mu} {\Lambda^{-1}}^d_b-(\partial_{\mu} \Lambda^{a}_{c}) {\Lambda^{-1}}^{c}_{b}$$
so that the spacetime connection coefficients stay the same. The spin connection which was vanishing would then take all possible forms of $-(\partial_{\mu} \Lambda^{a}_{c}) {\Lambda^{-1}}^{c}_{b}$. Vice versa, any flat spin connection can be brought to the zero value, at least locally. We would ignore global questions of possible extra topological freedom of cohomology type, therefore assuming the covariant model as just a covariantization of the purely Weitzenb{\" o}ck case.

Adding this spin connection does not change anything in the case of TEGR since it enters only via a surface term. However, for $f(T)$ the spin connection plays a less trivial role. Its variation in the action, if done in the purely inertial class,  gives the new equation \cite{Golovnev:2017dox}
\begin{equation}
\label{aseom}
f_{TT}\left(T_{\alpha\mu\nu}+g_{\alpha\mu}T_{\nu}-g_{\alpha\nu}T_{\mu}\right)\partial^{\alpha}{\mathbb T}=0.
\end{equation}
However, it is nothing but the antisymmetric part of equations for the tetrad:
\begin{equation}
\label{eom}
f_{T}\mathop{G_{\mu\nu}}\limits^{(0)}+f_{TT}S_{\mu\nu\alpha}\partial^{\alpha}{\mathbb T}+\frac12 \left(f-f_{T}{\mathbb T}\right)g_{\mu\nu}=8\pi G\cdot\Theta_{\mu\nu}
\end{equation}
where we have used the assumption that matter does not interact with the spin connection, its energy momentum tensor $\Theta_{\mu\nu}$ is symmetric, and also the simple relation $S_{\mu\nu\alpha}-S_{\nu\mu\alpha}=T_{\alpha\mu\nu}+g_{\alpha\mu}T_{\nu}-g_{\alpha\nu}T_{\mu}$.

This is an elementary and quite general fact for modified teleparallel gravities, that the flat spin connection of the covariantized version leads to the same equation as the antisymmetric part of the equation for the tetrad. Indeed, by construction the model is locally Lorentz invariant which means that the (Lorenzian) spin connection variation and the appropriate antisymmetric variation of the tetrad cancel each other in the variation of the action.

Therefore, the introduction of spin connection does not change the physical content of the theory. The pure tetrad formalism has 16 components of the tetrad as variables of the theory. Four of these components can be treated as purely gauge freedom, due to diffeomorphism invariance. In the case of TEGR there are six more components that are gauge freedom as well, due to the local Lorentz (pseudo)invariance. These considerations leave six physical variables, as in GR (two dynamical and four constrained but not arbitrary). In $f(T)$ theory, the local Lorentz symmetry is lost, at least partially, leaving more variables as physical. Formally, the invariance can be restored by introducing the flat spin connection. It represents six new variables, however without changing the number of the physical ones. The theory is now invariant under a simultaneous local Lorentz transformation of both the tetrad and the spin connection, and therefore six combinations of them are pure gauge. Indeed, one possible gauge choice is to set $\omega$ to zero.

Of course, this covariantization does not bring any new properties by itself. This is nothing but rewriting something in a way that allows a change of the description of the physical content. For example, the problems of bodies falling down on the Earth can also be written in a formally rotationally invariant way, if instead of fixing the vertical direction to the $z$ axis we formulate it along an explicitly introduced vector which gets its components changed (instead of only the $z$ one) if the Cartesian coordinates are rotated in a non-horizontal plane.

The antisymmetric part of the equations can also be thought of as regarding to the Lorentzian degrees of freedom. Indeed, Eqs. (\ref{eom}) and (\ref{aseom}) contain both the tetrad and the spin connection. However, the spin connection enters the torsion tensor components without any derivatives. Therefore, the (first) derivatives of the spin connection components would enter the equations only via the gradient of $\mathbb T$. Since the torsion scalar is a quadratic expression in terms of the connection coefficients (\ref{fullsp}), its dependence on the spin connection has a quadratic piece and a linear piece multiplied by derivatives of the tetrad.

To easily find the precise form of how the spin connection enters the equations of motion, we can compute $-2 \mathop\bigtriangledown\limits^{(0)}{\vphantom{\bigtriangledown}}_{\mu}T^{\mu}$ instead of $\mathbb T$. This is due to  Eq.\eqref{telepid} and the fact that the Levi-Civitian curvature does not depend on the spin connection at all. The torsion tensor is the antisymmetric part of the spacetime connection \eqref{fullsp}, and it has the term 
$e_a^{\alpha}(\omega^{a}{}_{b\mu}E^b_{\nu}-\omega^{a}{}_{b\nu}E^b_{\mu})$ with the spin connection, but for the torsion vector $T_{\mu}$ the spin connection will only enter as   $-e^{\alpha}_a\omega^a_{\hphantom{a}b\alpha}E^b_{\mu}$, due to its antisymmetry property. Therefore, we obtain that the contribution of the spin connection to the torsion vector divergence has the form 
\begin{equation} \label{divomega}
- \mathop\bigtriangledown\limits^{(0)}{\vphantom{\bigtriangledown}}_{\mu}(\omega^a_{\hphantom{a}b\alpha} e^{\alpha}_a E^b_{\nu}\eta^{cd}e^{\mu}_c e^{\nu}_d)=- (\mathop\bigtriangledown\limits^{(0)}{\vphantom{\bigtriangledown}}_{\mu}\omega^a_{\hphantom{a}b\alpha}) e^{\alpha}_a E^b_{\nu}\eta^{cd}e^{\mu}_c e^{\nu}_d-\omega^a_{\hphantom{a}b\alpha}\mathop\bigtriangledown\limits^{(0)}{\vphantom{\bigtriangledown}}_{\mu}(e^{\alpha}_aE^b_{\nu}\eta^{cd}e^{\mu}_c e^{\nu}_d).    
\end{equation}
The second term on the right hand side gives the linear in spin connection piece of $\mathbb T$. The first term has to be rewritten in such a way as to get rid of the derivatives of $\omega$, for it represents the dependence of $\mathbb T$ on the spin connection (which does not involve derivatives of $\omega$). It would actually give the quadratic in spin connection contribution to the torsion scalar. Indeed, the first term is 
\begin{equation}
- (\mathop\bigtriangledown\limits^{(0)}{\vphantom{\bigtriangledown}}_{\mu}\omega^{ab}{}_{\alpha})e^{\alpha}_a e^{\mu}_b,
\end{equation}
which consists of a term antisymmetric in $\alpha$ and $\mu$ derivatives of $\omega$ (due to the spin connection antisymmetry in $a$ and $b$, while the symmetry of Levi-Civita connection shows that the antisymmetrized covariant derivative coincides with the antisymmetrized partial derivative), and since the spin connection is flat, we have
\begin{equation}
    \partial_{\mu}\omega^a_{\hphantom{a}b\alpha }-\partial_{\alpha}\omega^a_{\hphantom{a}b\mu }=\omega^a_{\hphantom{a}c\alpha }\omega^c_{\hphantom{a}b\mu }-\omega^a_{\hphantom{a}c\mu }\omega^c_{\hphantom{a}b\alpha }
\end{equation}
from vanishing of its curvature tensor. Finally, the first term in \eqref{divomega} corresponds to
\begin{equation}
\left(\omega^a_{\hphantom{a}c\alpha }\omega^c_{\hphantom{a}b\mu }-\omega^a_{\hphantom{a}c\mu }\omega^c_{\hphantom{a}b\alpha } \right) e^{\alpha}_a e^{\mu}_b.
\end{equation}

The formulae above give the terms in $\mathbb T$ depending on the spin connection, and therefore the terms with the first derivatives of the spin connection in the gradient of it. Since in our teleparallel model the spin connection can be written in terms of arbitrary Lorentz matrices as  $\omega^a{}_{b\mu} = -\partial_{\mu} \Lambda^{a}_{c} \left(\Lambda^{-1}\right)^{c}_{b}$, we see the terms in the gradient of $\mathbb T$ that contain second order derivatives of the Lorentz matrices. Therefore, in the  antisymmetric equation $A^{\mu\nu\alpha}\partial_{\alpha}\mathbb T=0$,  second derivatives of Lorentz matrices enter revealing that such equations exhibit some novel dynamics in the Lorentzian sector.

The breakdown of the local Lorentz invariance means that, despite the equations of motion remaining second order in time derivatives (unlike $f(R)$ gravity), the number of degrees of freedom should be increased with respect to GR due to dynamics appearing in the sector of Lorentz rotations. In the covariant formulation, the local Lorentz symmetry is present, however new variables are introduced.

The Lorentz invariance violation of this sort is not extremely dangerous for the theory as long as the matter couples only to the metric, since such violation is not directly observable then. However quantum effects can in principle transport Lorentz violations from far remote sectors to observable phenomena \cite{AstrQ}. It is unclear how bad can it be in $f(\mathbb T)$ gravity case, since works on quantum modified teleparallel gravity are very  scarce. 

Also an interesting problem to analyze is the dynamics of fermions in teleparallel spacetimes. The only known reasonable way of coupling them is to the spin connection which corresponds to the Levi-Civita spacetime connection, however this would look quite inelegant from the fundamental teleparallel point of view. Covariant formulation can potentially give us some new angles for investigating modified procedures of matter coupling.

\section{Linear perturbations}

The weak gravity regime (understood as small perturbations around Minkowski spacetime with the trivial choice of tetrad) of $f(T)$ is (linearly) equivalent to GR, which in particular means that it exhibits the usual two polarizations of gravitons propagating at the speed of light  \cite{ Bamba:2013ooa,Cai:2018rzd}. 

This is a very elementary fact for perturbations around the trivial Minkowski tetrad $E^a_{\mu}=\delta^a_{\mu}$. Indeed, the perturbation of $\mathbb T$ is quadratic in derivatives of $\delta E$, and therefore the quadratic action is linear in $\delta \mathbb T$, which just gives the same result as in TEGR, and therefore as in GR.

The absence of new dynamical modes in $f(T)$ is also true for perturbations around cosmological solutions described by the simplest background tetrad $E^a_{\mu}=a(t)\cdot \delta^a_{\mu}$ \cite{Li:2011wu,Golovnev:2018wbh}. An explanation for this fact is unclear, unlike for the disappearance of new modes around Minkowski space. For cosmological perturbations, the calculations show that one of six new modes completely drops off from the linear equations, and the other five modes get constrained, not dynamical \cite{Golovnev:2018wbh}. As expected, the same perturbation analysis can be done in the covariant version of the theory \cite{mespin}.

An important issue about perturbation analysis in $f(T)$ gravity is that it is not enough to reproduce the metric perturbations, but one should also consider all Lorentzian perturbations in the tetrad field. The most general way to consider cosmological perturbations is as (see \cite{Golovnev:2018wbh} for definitions)
\begin{equation}
\begin{split}
E^0_0 & =  a(\tau)\cdot\left(1+\phi\right),\\
E^0_i & =  a(\tau)\cdot\left( \partial_i \beta+u_i\right),\\
E^a_0 & =  a(\tau)\cdot\left( \partial_a \zeta+v_a\right),\\
E^a_j & =  a(\tau)\cdot \left((1-\psi)\delta^a_j+\partial^2_{aj}\sigma+\epsilon_{ajk}\partial_k s+\partial_j c_a+\epsilon_{ajk}w_k+\frac12 h_{aj}\right),
\end{split}
\end{equation}
with the usual separation into scalar, vector and tensor perturbations. Taking only one particular tetrad choice for the metric perturbations instead of this full set leads to the wrong conclusion that perturbations are not possible at all, since an overdetermined system of equations is obtained \cite{CDDS}.

Having the same number of dynamical modes as in GR makes cosmological analysis simpler. Changes in the background dynamics can be used for attempts of solving such modern cosmology problems as the $H_0$ tension \cite{HHGZ} without running into contradictions with other data. Moreover, the theory predicts non-zero gravitational slip \cite{Golovnev:2018wbh} which can be used as a new and potentially observable feature of these alternatives to the standard cosmological model. On the other hand, it is also bad news for cosmology since $f(T)$ definitely has new dynamical modes, and therefore their absence in linear cosmological perturbations signals the strong coupling problem.

There are several ways to prove the presence of new modes. The most fundamental one is the Hamiltonian analysis, see the next section. On the other hand, one can either go to higher orders around simple backgrounds or study the perturbations around non-trivial backgrounds. Once considering this, we observe that there is compelling evidence of existence of new modes in $f(T)$ gravity after all.

Nonlinear look has been taken in Ref. \cite{Jimenez:2020ofm} where the action around the trivial Minkowski background was taken up to the fourth order which exhibited a new dynamical mode there. On the other hand, linear perturbations about other Minkowski backgrounds (with boosted and rotated background tetrads) were considered in  \cite{wenew}, and new equations signaling a new mode, but with rather unusual properties, were found.

It is also possible to approach such questions indirectly, by studying a class of solutions instead of really counting the number of degrees of freedom. For example, in the  \cite{Finch:2018gkh} a model with $f({\mathbb T})={\mathbb T}+\alpha {\mathbb T}^n$ was used to reproduce galactic rotation curves. It was found that in the $n \to 1$ limit the results were not approaching the case of GR. Since the model with the limiting Lagrangian is then equivalent to GR, it as an indication of new degree(s) of freedom having some non-trivial dynamics. 

In \cite{noFinch}, it was claimed that  \cite{Finch:2018gkh} was based on wrong considerations of an earlier paper \cite{RugRad}, in terms of a ``bad'' tetrad. However, the authors of \cite{RugRad} actually did check the antisymmetric part of equations, and their tetrad is different from the tetrad of \cite{noFinch} only by the sign of components which do not contain unknown functions of $r$. Since changing the overall sign is a global transformation which changes nothing, it means that this is basically the same tetrad just with change of signs of unknown functions (to be found by solving the symmetric part of equations). Therefore, this is the same ``good'' tetrad, though their construction with positive values of the functions might present something very new and not close to anything from GR. It would be good to accurately study all possible options. This is a very interesting issue, not only for substituting Dark Matter with modified gravity, but also for studying the fundamental properties of $f(T)$ gravity.

\section{Hamiltonian formalism and remnant symmetry}

There are at least three different Hamiltonian analyses of $f(T)$ gravity in the current literature. Although some contradict to each other, all agree that some new degree(s) of freedom do(es) appear.

It first appeared in \cite{Li:2011rn}. They claim three new degrees of freedom in $(3+1)-$ dimensional space. The analysis there is not very accurate, and in many parts it seems to be  that the only outcome is that all Lorentz constraints become second class and nothing else. This would mean full break down of the local Lorentz symmetry, which is not true. So far, it is known that there are physically interesting backgrounds with no new dynamical modes in linear perturbations, and that constant $\mathbb T$ cases are somewhat special. Note also that their claim on the second class nature of Lorentz constraints is related to the unique determination of Lagrange multipliers, which  they argue about in the Appendix. However, calculations are not fully rigorous, since in their derivations they use some ``calculation method'' where it is assumed that the metric is diagonal, which is not justified at all.

The new impetus to the field was given by  \cite{Ferraro:2018tpu,Ferraro:2018axk,Ferraro:2020tqk} which claim existence of only one new dynamical mode. Unfortunately, as was noticed in  \cite{Blagoj}, the Poisson brackets of the Lorentz constraints are incomplete, therefore the main conclusion is wrong. To be more precise, the piece which depends on spatially non-constant $\mathbb T$ (or spatially non-constant auxiliary field $\phi$) is missed. However, for this particular case their analysis seems to be consistent, therefore it can still give information about spatially constant $\mathbb T$ solutions, which corresponds to the important Minkowski and FLRW cases.

Finally, the last analysis came in \cite{Blagoj}. It also has problems. The local Lorentz algebra is said to be broken down, however their matrix of Poisson brackets of Lorentz constraints $C_{ij}$ and $C_{mn}$ has a non-trivial term given by $\phi_{,k}\delta^{0kp}_{ijm}g_{pn}$ where $\delta$ is the generalized Kronecker symbol (we abandoned unnecessary complications of the covariant approach of Dirac Hamiltonian analysis). It does not seem to be non-degenerate. Right after writing out the Poisson bracket the authors claim that the presence of a non-trivial term shows that the constraints are second class. But to really prove it just from this bracket it would be necessary to check that the $6\times 6$ matrix of this term is non-degenerate. Otherwise, it is a combination of first and second class constraints.

Also they claim no new modes around any ${\mathbb T}= \text{const}$ solution which contradicts simple calculations around boosted or rotated trivial tetrads \cite{wenew}. Moreover, they say it makes sense since the model is then equivalent to GR. This is incorrect. Only the background equation for a ${\mathbb T}=\text{const}$ solution is the same as in GR with a cosmological constant. Dynamics of perturbations, even the linear ones, is generically not equivalent.
 
All the above analyses were done in the pure tetrad formulation. However, since the covariant formulation is equivalent to it, there should be no difference. One can indeed find direct support for this claim \cite{BHP}, however it does not solve the problems which are still there in the Hamiltonian analysis of these models.

Getting a clear, reliable and understandable Hamiltonian analysis is very important. A first step in this direction could consist of getting a homogeneous Hamiltonian description of gravity theories based on the teleparallel formalism \cite{Blixt:2020ekl}. Moreover, the question of how much of the local Lorentz symmetry is violated is very important for this endeavor. This is known under the name of remnant symmetries and has already been considered \cite{Ferraro:2014owa}. However there is still a lot left to study.

A big part of \cite{Ferraro:2014owa} is devoted to infinitesimal transformations which leave $\mathbb T$ invariant at the linear level. The relevance of it is not very clear. For the linear invariance of perturbation equations it is also necessary to have invariance of the superpotential, or invariance of the action to quadratic order. For example, in linear cosmological perturbations \cite{Golovnev:2018wbh} only one of the Lorentz variables (the pseudoscalar perturbation $s$) does not show up in the linearized equations while the set of linearized $\mathbb T$ invariance is wider \cite{Ferraro:2014owa}.

All transformations which keep $\mathbb T$ exactly unchanged is not precisely the same as the symmetry group of the theory. First, there are different amounts of symmetries for different solutions, which means that the number of degrees of freedom might indeed be not well-defined globally. Second, these sets of invariance are not groups. For example, around the trivial Minkoswki space those sets contain two-dimensional Abelian subgroups of the Lorentz group which correspond to the rotated and boosted tetrads of \cite{wenew}, but one cannot take two elements from different Abelian subgroups. A better understanding of symmetry properties will be very important for doing a good Hamiltonian analysis.

\section{Causality and Cauchy development}

There are also other strategies to study foundational issues of modified teleparallel gravity. Not only the number of degrees of freedom is important but also their physical nature and properties. 

In \cite{Ong:2013qja}, acausal propagation was claimed to be highly likely in $f(T)$ gravity by using the method of characteristics, however the conclusions rely on the results on degrees of freedom from the first Hamiltonian analysis \cite{Li:2011rn}. There could definitely be problems with the theory with Poisson brackets vanishing at some points in the configuration space, since some solutions apparently have more remnant symmetry than others. The types of derivatives (time or space) in the equations of motion can indeed vary depending on the boosted tetrad background chosen for the Minkowski spacetime \cite{wenew}. 

Another claim from the same paper \cite{Ong:2013qja} is absence of predictable Cauchy development. For that they present a family of spatially flat FRW cosmological solutions, with a time-dependent boost of the simple standard tetrad in a fixed direction. Indeed, the freedom of choosing the time dependence of the boost makes the torsion tensor unpredictable.

The discussion on the unpredictability of the torsion tensor starts from the paper of Kopczy\'nski on new general relativity (NGR)  \cite{Kopczynski:1982}, where the author used perturbation of the spin connection instead of the Lorentz rotation of the tetrad. Nester in his paper \cite{Nester:1988} claims that Kopczy\'nski did not require that the spin connection had to remain flat, and he adds an  integrability condition that the variation of the spin connection can be represented in terms of the Lorentz matrices, as it should be. After that, Nester's conclusion (justified with words like ``presumably'') is that the predictability is spoiled only for very specific solutions (though very physically interesting ones, we should say), however later in some papers on $f(T)$ \cite{Ong:2013qja,Chen:2014qtl} this  opinion on these issues seems to be radically changed without much explanation on that.

Note that the same unpredictability of the torsion tensor is true for TEGR, and the reason to not worry about this is simply that TEGR is a gauge theory with respect to the local Lorentz rotations of the tetrad, and only the metric is physical. Therefore, we do not care about the value of the torsion tensor as long as we can predict the metric. We might also not care in $f(T)$. But that would require a clear answer to the question on what part of the freedom can indeed be regarded as a genuinely surviving gauge symmetry. And when it is not, we see that perturbations can behave differently around different tetrad backgrounds \cite{wenew}. However, once we have chosen a particular background, can we build a good perturbation theory having some well-defined remnant gauge symmetry?

The minimal physical requirement would be to have the metric predictable. Most of the known unpredictability examples do not directly address this issue since the Lorentz transformations do not change the metric. A stronger statement can be found in the paper \cite{Izumi:2013dca} where the authors generalized the case of time-dependent boost of the tetrad to the Bianchi I cosmology. It turned out that the difference of two Hubble rates is unpredictable which already brings the issue to the metric sector. Their text can be read as if they blame diffeomorphism invariance for that, which is not correct. Apparently, the reason is that their boost parameter $\theta(t)$ is a new function which does not have a new equation due to the fact that it is not restricted by antisymmetric equations.

However, as we previously discussed, even with background metric unchanged, perturbations can behave differently around different backgrounds with the same metric. In \cite{wenew}, we considered a special case of a Lorentz boost of the trivial Minkowski tetrad along the $x$ direction with the boost parameter depending only on $z$. The linear theory tells that $\mathbb T$ does not depend on $t$ and $x$ but can have arbitrary dependence on $y$ and $z$. With different dependence of parameters on spacetime coordinates, one can change the sets of coordinates on which $\mathbb T$ can or cannot depend.

What would happen in such cases at higher orders? Without a full-fledged gauge symmetry, the unpredictability of the torsion tensor can hardly stay away from the metric sector, but logically there are two possibilities. Either the higher orders would finally require $\mathbb T=\text{const}$, and then the purely linearized treatment cannot be considered as anything reliable, but at least the torsion scalar would be almost predictable, up to the arbitrary constant. Or this kind of freedom remains at the second order and beyond too, and then the second order correction to the symmetric part would have unpredictable influence on the gravitons via the terms like $f_T\mathop{G_{\mu\nu}}\limits^{(0)}$ (and the term proportional to the metric) which would have unknown dependence of scalar coefficients on the spacetime point. Either option looks unpleasant but from the theoretical viewpoint it anyway seems very interesting to better study the foundational properties of $f(T)$ gravity. 

\section{Discussion}

In the recent years, $f(T)$ gravity theories are often used for cosmological model building, sometimes quite successfully \cite{HHGZ}. Also, there are some (not so common) attempts to smooth GR singularities in the Big Bang and Black Holes. Despite this versatility and success of it as a theoretical playground, there are many points that are still not very well understood concerning the degrees of freedom and different viewpoints on the breaking of Lorentz invariance.

Up to now, many research pieces that can be found in the literature have to be interpreted with great care. The local Lorentz invariance is broken (in pure tetrad formulation) with details depending on the chosen tetrad background. The set of remnant symmetries (which does not form a group) varies for different tetrad configurations, which give different linear perturbation properties even when treating the same metric. There are reasons to believe that the theory has a highly non-linear nature \cite{Ong:2013qja} so that background solutions and linear perturbations might be misleading about the true non-perturbative behavior. Therefore, although a very challenging task, a deeper investigation is still necessary.

For many modified gravity models, some transformations of variables allow to get a simpler picture. However, finding a real Einstein frame for $f(T)$ would have required a way to encode its Lorentz violation in some extra fields. Therefore a simple conformal transformation cannot do it, as has been noticed already some time ago \cite{Yang:2010ji,Wright:2016ayu}. Trying to move the factor of $\phi$ inside the torsion scalar in the Lagrangian term of $\phi\mathbb T$ produces the new Lorentz-breaking term with $T^{\mu}\partial_{\mu}\phi$. Disformal transformations did not help either \cite{wedis}. A disformal transformation, with coefficients depending only on the scalar field, changes the coefficient in front of $ET^{\mu}\partial_{\mu}\phi$ action term only by the factor of $C^2$, the conformal coefficient of the transformation \cite{wedis}. 

Another interesting opportunity is to look for other non-trivial solutions using the method of null tetrads \cite{Bejarano:2014bca,Nashed:2014yea,Bejarano:2017akj}. One can, for example, construct a ${\mathbb T}=0$ cosmological solution \cite{Bejarano:2017akj}. It can be a starting point for studying cosmological perturbations around this symmetry breaking background and possibly with dynamical new mode(s) even at the linear order.

As we have seen, there are many directions to be taken for a better understanding of $f(T)$ gravity. Even if the model has a very high risk of being pathological, it is still worth to study its properties and problems with more detail. Cosmological and other applications seem promising, at least at the naive level of not worrying much about the nonlinear problems. It is clear that a good understanding of $f(T)$ gravity is important and will lay the foundations for studying generalizations to different models such as new general relativity, $f(T,B)$ gravity, symmetric teleparallel frameworks and other options that share similar properties. 

\section*{Acknowledgments}

M. J. G. was funded by CONICYT-FONDECYT postdoctoral grant 3190531.

\end{document}